\documentclass[12pt]{article} 
\usepackage{graphicx,floatflt,amssymb,rotate,epsfig,subfig}
\usepackage{mathrsfs}
\usepackage{amssymb}
\usepackage{amsmath}
\usepackage{color}
\usepackage{graphicx}
\usepackage{subfig}
\usepackage{multirow}
\usepackage{cite}
\usepackage{float}
\usepackage{pstricks}
\usepackage[toc,page]{appendix}
\usepackage{pifont}
\usepackage{braket}
\usepackage{bbold}
\usepackage{slashed}
\usepackage[normalem]{ulem}
\usepackage[colorlinks=true,
linkcolor=red,
urlcolor=blue,
citecolor=blue]{hyperref}
\usepackage{color}

\def\Bbar{\overline{B}}

\def\nubar{{\overline{\nu}}}

\def\Re{\mathcal{R}e}

\def\Dst{{D^*}}

\usepackage{relsize}
\def\Babar{{\mbox{\slshape B\kern-0.1em{\smaller A}\kern-0.1em B\kern-0.1em{\smaller A\kern-0.2em R}}}}

\usepackage{enumitem}

\newcommand{\ba}{\begin{array}}
\newcommand{\ea}{\end{array}}
\def\beq{\begin{equation}}
\def\eeq{\end{equation}}
\def\bea{\begin{eqnarray}}
\def\eea{\end{eqnarray}}

\def\roughly#1{\mathrel{\raise.3ex\hbox
{$#1$\kern-.75em\lower1ex\hbox{$\sim$}}}}

\def\sla#1{\raise.15ex\hbox{$/$}\kern-.57em #1}

\def\bd{B_d^0}

\def\order{\lower 1.8ex \hbox{\LARGE\~{}}}

\def\mET{E_T \hspace{-1.0em}/\;\:}

\def\bd0tau{B\to D \tau\nu_{\tau}}

\def\be {\begin{equation}}
\def\ee {\end{equation}}

 \usepackage[normalem]{ulem}

 \definecolor{darkgreen}{cmyk}{1,0,1,0.4}
 
 \definecolor{pink}{cmyk}{0.4,1,0.3,0}

\def\com2#1{\textcolor{red}{\it{#1}}}

\newcommand{\mptvec}{ \, \not \! \vec{P}_T}

\begin{document}
\vskip 30pt 

\begin{center}  
{\Large \bf Collider signature of $U_1$ Leptoquark and constraints from $b\to c$ observables} \\
\vspace*{1cm}  
\renewcommand{\thefootnote}{\fnsymbol{footnote}}  
{{\sf Aritra Biswas$^1$\footnote{email: iluvnpur@gmail.com}}, 
{\sf ~Dilip Kumar Ghosh$^1$\footnote{email: tpdkg@iacs.res.in}}, 
{\sf ~Nivedita Ghosh$^1$\footnote{email: tpng@iacs.res.in}}, 
{\sf Avirup Shaw$^1$\footnote{email: avirup.cu@gmail.com}}, 
{\sf Abhaya Kumar Swain$^1$\footnote{email: abhayakumarswain53@gmail.com }}
}\\  
\vspace{10pt}  
{ {\em $^1$School of Physical Sciences, Indian Association for the Cultivation of Science,\\
2A $\&$ 2B Raja S.C. Mullick Road, Jadavpur, Kolkata 700 032, India}
}
\normalsize  
\end{center} 

\begin{abstract}  
\noindent 
One of the most popular models that is known to be able to solve the lepton flavour universality violating charged ($b\to c$) and neutral current ($b\to s$) anomalies is the Leptoquark Model. In this work we examine the {\it multijet} + $\mET$ collider signature of a vector leptoquark ($U_1$) which has the potential to mediate both the charged and neutral current processes at tree level. From our collider analysis we derive the exclusion limits on mass for the $U_1$ leptoquark at 95\% C.L. at the current and future experiment of the Large Hadron Collider. We also calculate the effect of such a leptoquark in $b\to c$ observables.
\vskip 5pt \noindent  
\end{abstract}

\renewcommand{\thesection}{\Roman{section}}  
\setcounter{footnote}{0}  
\renewcommand{\thefootnote}{\arabic{footnote}}

\section{Introduction}\label{intro}
The Standard Model (SM) of particle physics is the most successful theoretical description of the experimentally detected spectrum of fundamental particles till date. This description is based on the gauge invariance of the local group ${\rm {SU(3)_C\times SU(2)_L\times U(1)_Y}}$. The quarks and leptons enter this description as independent fields. However, the success of the SM as a quantum field theory is crucially dependent on the cancellation between the lepton and quark contributions to triangle anomalies of gauged currents. As such, it is only logical to expect that a more fundamental description of these particles might incorporate an interrelation between the quarks and the leptons~\cite{Buchmuller:1986zs}.

The Leptoquark (or Lepto-quark) (LQ) is such an extension of the SM where the LQs are hypothetical particles which mediate interactions between quarks and leptons at tree-level. Such scenarios emerge naturally in several extensions of the SM~(e.g., composite models \cite{Schrempp:1984nj}, Grand Unified Theories \cite{Georgi:1974sy, Pati:1973rp, Dimopoulos:1979es, Dimopoulos:1979sp, Langacker:1980js, Senjanovic:1982ex, Cashmore:1985xn, Pati:1974yy}, superstring-inspired ${\rm E_6}$ models \cite{Green:1984sg, Witten:1985xc, Gross:1984dd, Hewett:1988xc} etc). The discovery of LQs would thus be a signal for matter unification. As such, these particles have extensively been discussed theoretically for over forty years, both from the point of view of their diverse phenomenological aspects~\cite{Davidson:1993qk,Hewett:1997ce,Nath:2006ut}, and specific properties~\cite{Shanker:1981mj,Shanker:1982nd,Buchmuller:1986iq,Buchmuller:1986zs,Hewett:1987yg,Leurer:1993em,Leurer:1993qx, Dorsner:2014axa,Allanach:2015ria,Evans:2015ita, Li:2016vvp, Diaz:2017lit,Dumont:2016xpj,Faroughy:2016osc,Greljo:2017vvb,Dorsner:2017ufx,Allanach:2017bta,Crivellin:2017zlb,Hiller:2017bzc,Buttazzo:2017ixm,Calibbi:2017qbu,Sahoo:2016pet, Altmannshofer:2017poe}. A considerable amount of work regarding LQs has also been undertaken from the experimental side. However, the major part of these searches have been directed towards scalar LQs~\cite{Baumgartel:2014dqa,Aad:2015caa,Aaboud:2016qeg,Sirunyan:2017yrk,Sirunyan:2018nkj}. Experimental studies on vector LQs, though present in the literature~\cite{Aaltonen:2007rb, Sirunyan:2018ruf, Sirunyan:2018kzh}, are scarce in number mainly because of some additional model dependent parameters. 

On a different note, there have been constant and consistent hints towards the presence of lepton flavour universality violating (LFUV) new physics (NP) both in charged-current~\cite{Lees:2012xj,Lees:2013uzd,Aaij:2015yra,Huschle:2015rga,Sato:2016svk,Hirose:2016wfn} and neutral-current~\cite{Aaij:2014ora,Aaij:2015oid,Aaij:2017vbb} processes over the last few years. These flavour anomalies exhibit diverse phenomenological roles in validating/invalidating or constraining a plethora of existing NP models. Various versions of LQ models have also been used in explaining these anomalies~\cite{Sakaki:2013bfa, Popov:2016fzr, Chen:2017hir, Alok:2017sui, Crivellin:2017zlb, Assad:2017iib, Aloni:2017ixa,Wold:2017wdj,Muller:2018nwq,Hiller:2018wbv,Biswas:2018jun,Fajfer:2018bfj,Monteux:2018ufc, Kumar:2018kmr, Crivellin:2018yvo}. The advantage in doing so is that LQ is one of those few models which allows for all the different kinds of NP interactions (based on their Lorentz structures, viz. scalar, pseudo-scalar, vector, axial-vector and tensor) that have the potential to explain such deviations. If LQs are potential candidates for explaining such anomalies, it is imperative that one carefully investigates the production and decay signatures of these entities at collider experiments and predict observables which help in their detection. As a result, the phenomenological community has recently displayed a lot of interest in collider studies of LQs~\cite{Dorsner:2014axa,Allanach:2015ria,Evans:2015ita,Diaz:2017lit,Greljo:2017vvb, Dorsner:2017ufx, Bandyopadhyay:2018syt, Vignaroli:2018lpq, Aaltonen:2007rb, Dorsner:2018ynv, Sirunyan:2018ruf, Biswas:2018iak}. However, collider searches dedicated to vector LQs in particular are very limited in the literature \cite{Aaltonen:2007rb, Diaz:2017lit, Biswas:2018iak, Dorsner:2018ynv, Sirunyan:2018ruf}. 

In the present article,  we choose a particular vector LQ $U_1$ which is among those few LQs that, in the most general case (i.e., with couplings to all three generations of fermion) has the potential to explain the data of both neutral current (e.g., $\mathcal{R}_{K^{(*)}}$) and the charged current (e.g., $\mathcal{R}_{D^{(*)}}$) anomalies.
However, the most recent data on $\mathcal{R}_{D^{(*)}}$ due to Belle \cite{Abdesselam:2019dgh} has brought the global deviation for these ratios with respect to SM down from $\sim 4\sigma$ to $\sim 3.1\sigma$. This, in principle, should result in even tighter constraints for the allowed parameter space corresponding to the $U_1$ vector LQ model. With this spirit, we propose a scenario where the $U_1$ vector LQ can couple with second and third generations of quarks but with only third generation of leptons. Therefore, the parameter space for the variant of the model discussed in the current article will not be affected by the $\mathcal{R}_{K^{(*)}}$ observables. However, the implications of the $\mathcal{R}_{D^{(*)}}$ ratios will result in a constrained parameter space, which has also been looked into in the scope of this article. Further, this LQ has baryon and lepton number conserving couplings. Consequently, there is no possibility of proton decay being mediated by $U_1$\footnote{At this point we remark in passing that, the lepton and baryon number violating LQs are very heavy in order to avoid bounds from proton decay. However, the LQs with the baryon and lepton number conserving couplings restrict proton decay and could be light enough to be seen in the LHC~\cite{Dorsner:2016wpm}.}. On top of that, the NP interaction term that imparts to the $\mathcal{R}_{D^{(*)}}$ observable will also contribute to the production of di-top plus missing energy. In this set up we perform a comprehensive collider analysis of $U_1$ vector LQ via {\it multijet} + $\mET$ final states.  We have utilized several interesting kinematic variables which best exploit the available kinematic information between the signal and background events to maximize the collider reach for 13 TeV LHC. Our analysis shows that the $U_1$ vector LQ can be excluded up to 2020 (2230) GeV at 95\% C.L. for an integrated luminosity of 300 (3000) $\rm{fb}^{-1}$. After obtaining the bound on the mass of the $U_1$ vector LQ from the collider analysis, we hence incorporate the corresponding flavour analysis to provide a quantitative estimate of how well this present version of the $U_1$ vector LQ model describes the current $b\to c$ data.

The paper is organised as follows. We briefly describe the Lagrangian for the $U_1$ vector LQ and set our convention in section~\ref{u1model}. Section~\ref{collider} we perform the collider analysis for $U_1$ via {\it multijet} + $\mET$ final states. In section~\ref{flav} we study the $b\to c$ observables mediated by the $U_1$ vector LQ. Finally, we summarize our results in section \ref{concl}.

\section{Effective Lagrangian of $U_1$ vector Leptoquark}\label{u1model}
It has been already mentioned that, LQs are special particles that appear naturally in particular extensions of the SM. Depending on the considered model, the LQs may be scalar (spin  0) or vector (spin 1) particles. All the LQs are colour-triplet and carry both baryon as well as lepton numbers. As a consequence they are able to mediate transitions between the quark and lepton sectors. Apart from the SM particles, a general LQ model\footnote{For detailed discussions regarding LQ scenarios, one can  look into~\cite{Dorsner:2016wpm}.} contains at least two massive neutrinos and twelve LQ particles. Among the twelve LQs, six are scalars ($S_3, R_2, \tilde{R_2}, \tilde{S_1}, S_1, \bar{S_1}$) and the rest ($U_3, V_2, \tilde{V_2}, \tilde{U_1}, U_1, \bar{U_1}$) transform vectorially under Lorentz transformations. As discussed earlier, the focus for the rest of our article  will be on $U_1$ vector LQ. The Lagrangian for kinetic and mass terms of $U_1$ vector LQ is given by \cite{Dorsner:2018ynv}:
\begin{equation}
\mathcal{L}^\mathrm{1}_{U_1}= -\frac{1}{2} U_{\mu\nu}^{\dagger} U_{\mu\nu} - i g_s \kappa~ U_{1\mu}^{\dagger} T^a U_{1 \nu} G^a_{\mu \nu}+ m_{U_1}^2 U_{1\mu}^\dagger U_{1\mu}\;,
\label{eq:kin}
\end{equation}
where $m_{U_1}$ denotes the mass of $U_1$ vector LQ. $\kappa$ is a dimensionless coupling that depends on the ultraviolet origin of the vector \cite{Dorsner:2018ynv}. For the minimal coupling case $\kappa = 0$, while for the Yang-Mills case $\kappa =1$  \cite{Dorsner:2018ynv}. Throughout our analysis we have assumed $\kappa =1$. $U_{\mu\nu} = D_\mu  U_{1\nu} - D_\nu  U_{1\mu}$ is a field strength tensor. $G^a_{\mu\nu}$ is the gluon field strength tensor, $T^a$ is the Gell-Mann matrix and $g_s$ is the QCD coupling strength. Generically, the Yukawa Lagrangian of the $U_1$ with the SM fermion bilinear can be written as~\cite{Dorsner:2016wpm}:  \begin{equation}
\mathcal{L}^\mathrm{2}_{U_1}=\left(h_{1L}^{ij}\bar{Q}_{iL}\gamma^\mu L_{jL}+h_{1R}^{ij}\bar{d}_{iR}\gamma^\mu l_{jR}\right)U_{1\mu} + {\rm h.c.}\;\;.
\label{Lagrangianv2}
\end{equation}
The gauge quantum numbers for $U_1$ under the SM gauge group ${\rm {SU(3)_C\times SU(2)_L\times U(1)_Y}}$ are $({\bf{3,1}},\frac23)$. $Q^{\sf T}_L\equiv (u~~~d)$ denotes the left handed quark doublet, $L^{\sf T}_L\equiv (\nu_{l}~~~l)$ stands for the left handed lepton doublet, $d_R$ is the right handed down type quark singlet and $l_R$ represents the right handed charged lepton. $h^{ij}_{1L(R)}$ are the left (right) handed Yukawa coupling constants while $i, j\equiv 1,2,3$ specify the fermion generation indices.

\section{Collider analysis}\label{collider}
\begin{figure}[t]
\centering
\includegraphics[keepaspectratio=true,scale=0.3]{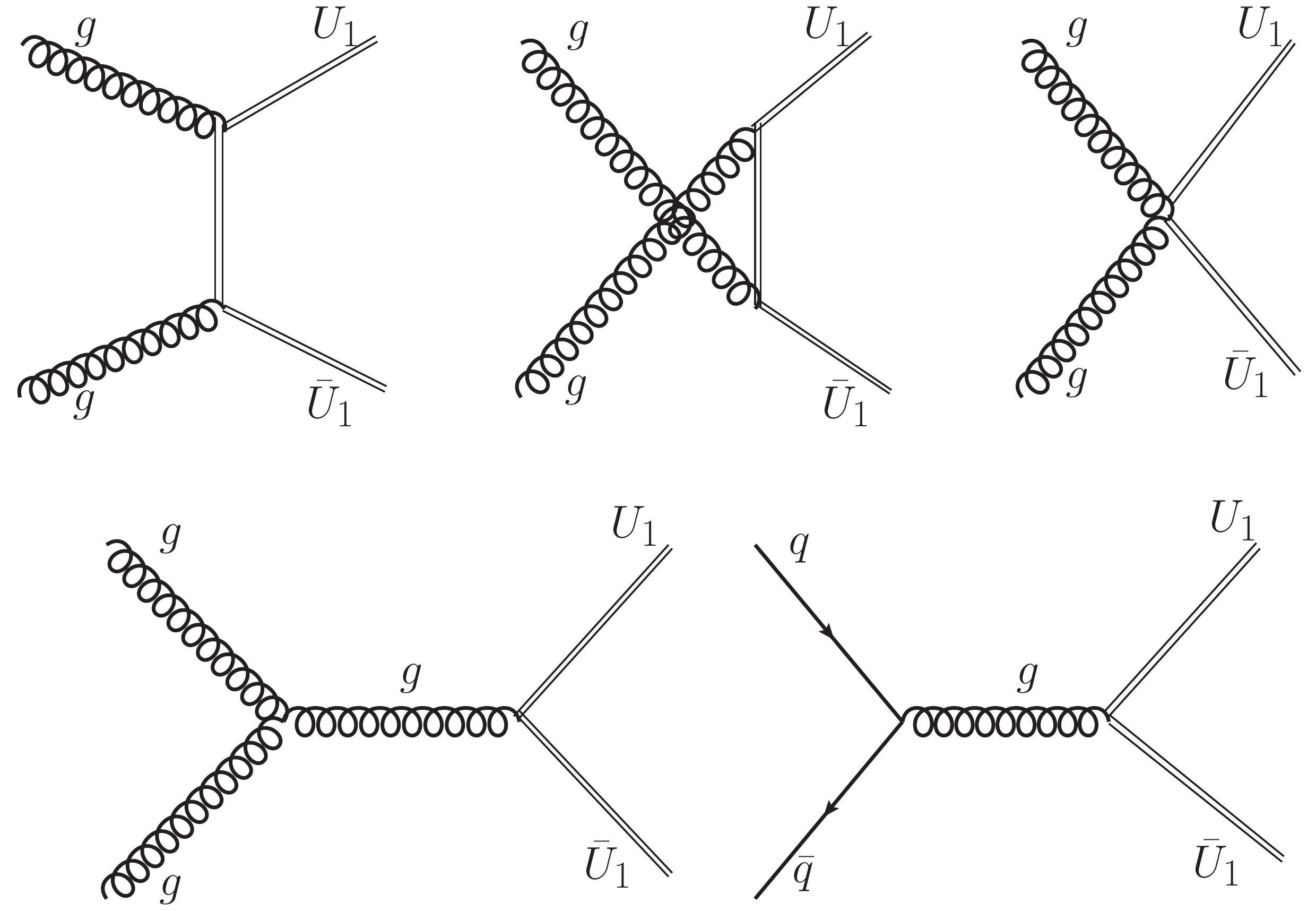}\hspace{2cm}
\includegraphics[keepaspectratio=true,scale=0.45]{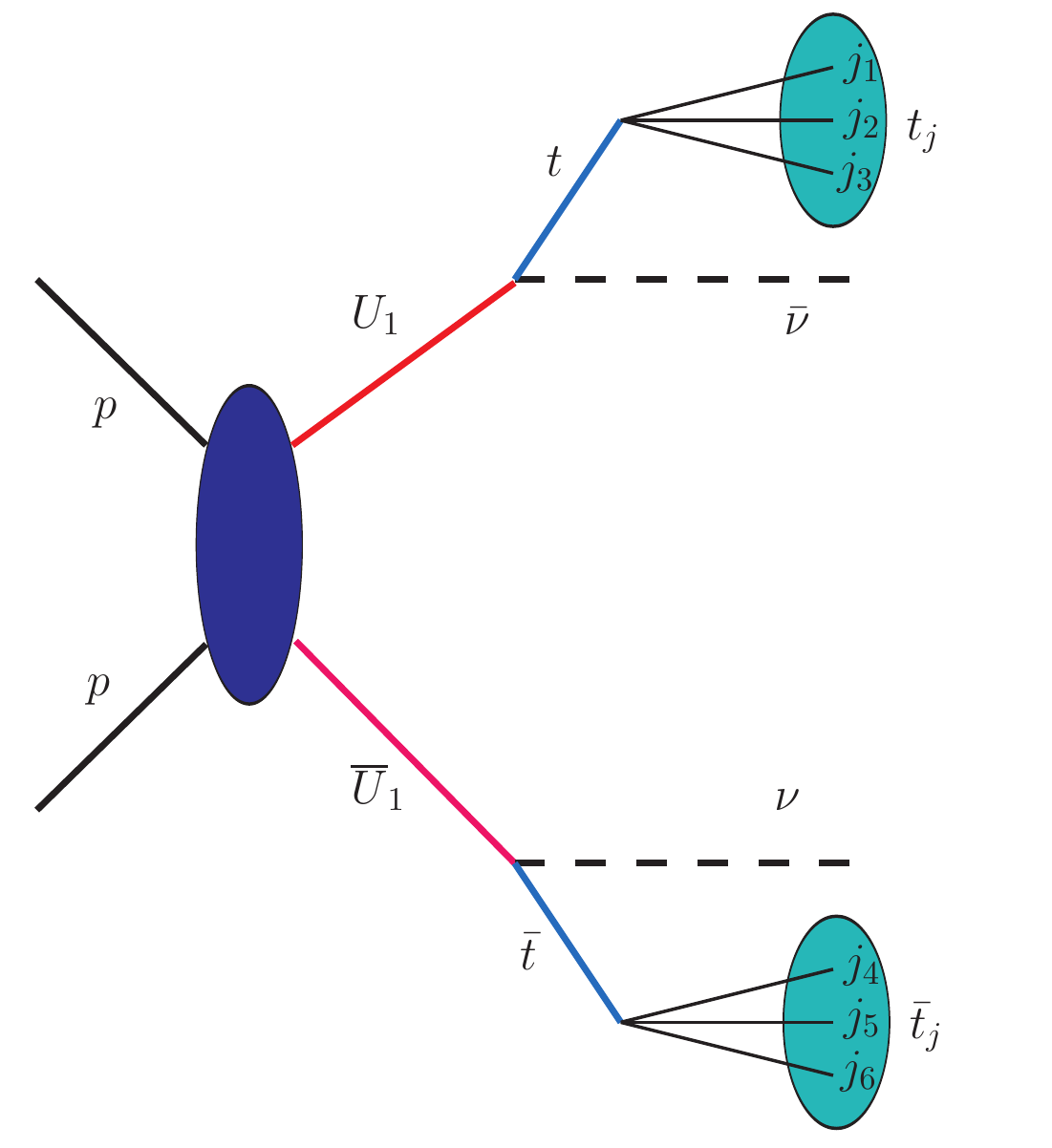}
\caption{(Left) panel displays the Feynman diagrams which contribute the $U_1$ vector LQ pair production. (Right) panel is a representative kinematic diagram for the pair production of the $U_1$ vector LQ, each of which subsequently decays to the top quark and neutrino. The top (anti-top) quark further decays hadronically leading to multijets and missing energy in the final state. Since the top (anti-top) is produced from the $U_1$ vector LQ, it is highly boosted and hence the decays are collimated. The collimated objects are diagrammatically represented as green blobs. The black dashed lines represent the neutrinos which are undetected in the detector and give rise to momentum imbalance, viz. the missing transverse momenta.}
\label{fig:di-toptopology}
\end{figure}

We begin our collider analysis by specifying the signal topology that we consider:
\begin{eqnarray}
 p\;p \to U_1 \overline{U_1} \to (t ~ \bar{\nu}) + (\bar{t} ~ \nu) \label{di-top_signal}\;\;,
\end{eqnarray}
where the $\overline{U_1}$ represents the anti-particle of the $U_1$ vector LQ. The interaction term responsible for the pair production of the $U_1$ vector LQ at the LHC is  given in the Lagrangian $\mathcal{L}^\mathrm{1}_{U_1}$ (see eq.~\ref{eq:kin}). The Feynman diagrams for the pair production of the $U_1$ vector LQ given in the left panel of fig.~\ref{fig:di-toptopology}. The signal in eq.~\ref{di-top_signal} is characterized by the presence of di-top and large missing energy  ($\slashed{E}_T$) in the final state which we designated as di-top signal. We have assumed here that the $U_1$ vector LQ couples only to the third generation quarks and leptons by assigning non-zero values to the corresponding couplings while setting the other couplings to zero. Furthermore, we also consider the coupling of $U_1$ to top quark and neutrino to be equal to that of the bottom and tau-lepton in order to simplify\footnote{However, a more general analysis can be done using different values for the couplings.  Such an analysis can potentially provide limits in coupling-mass plane. We, however, choose to make the  analysis as simple as possible by reducing free parameters of the model.} our analysis. Therefore, the branching ratio for each channel is approximately 50\%. Each top quark in the final state is assumed to decay hadronically. Since the top quark is produced from the (very heavy) $U_1$ vector LQ, it is expected to have large boost. The corresponding decays are hence collimated and fall inside a large radius jet (marked by green blobs in the right panel of fig.~\ref{fig:di-toptopology}) which is discussed below. Therefore, we expect at least two large radius jets, $N_{fatJet}$, with transverse momentum $(P_T) >$ 50 GeV and large missing transverse energy ($\mET > 100$ GeV). 
Since the signal has multijets in the final state, we further demand that there should at least be two jets in the final state with $P^j_T \ge 20$ GeV and $|\eta_j| \le 2.4$ and the reconstructed leptons (electrons and muons) with $P^l_T \ge 10$ GeV and $|\eta_l|\le 2.4$ are vetoed.

The SM processes which contribute as backgrounds to the above final state are $t\bar{t}$ + jets,  $V+$ jets, where $V = W^{\pm}, ~Z$ and QCD multijets (up to four jets). Since the QCD multijets have very small missing transverse energy, it can be handled using a moderate to large missing energy cut, so the dominant contribution comes from the top pair events and $V+$ jets backgrounds. The signal significance can be maximized subject to appropriate choices for the kinematic variables. The events corresponding to the signal and SM backgrounds in our analysis have been generated using {\tt Madgraph5}~\cite{Alwall:2014hca} with the {\tt NNPDF3.0} parton distribution functions~\cite{Ball:2014uwa}. The UFO model files required for the Madgraph analysis have been obtained from {\tt FeynRules}~\cite{Alloul:2013bka} after a proper implementation of the model. Following this parton \-level analysis, the parton showering and hadronisation are performed using {\tt Pythia}~\cite{Sjostrand:2006za}. We use {\tt Delphes}(v3)~\cite{deFavereau:2013fsa} for the corresponding detector level simulation after the showering/hadronisation. The jet construction at this level has been performed using fastjet~\cite{Cacciari:2011ma} which involves the anti-$K_{T}$ jet algorithm  with radius $R = 0.5$ and $P_T > 20$ GeV. The hard-jet background as well as the signal events have been properly matched using the MLM matching scheme~\cite{Hoche:2006ph}. The signal and backgrounds except $V +$jets are matched up to 2 jets and matching for $V +$jets are done up to 4 jets. After getting the reconstructed jets in each event, we again pass the jets through the fastjet with radius\footnote{Since the top will be highly boosted, the top decay products will fall in the large radius jets of radius of 0.8 on a statistical basis and we have also checked that changing the jet radius will have mild effect on the results presented here.} $R = 0.8$ to get the large radius jets with $P_T > 50$ GeV for the di-top signal.   

The cross section used in this analysis for the background process $t\bar{t}$ is 815.96 pb~\cite{Czakon:2011xx} as calculated with the Top++2.0 program to NNLO in perturbative QCD, with soft-gluon resummation to NNLL order assuming a top quark mass of 173.2 GeV. The single vector boson production cross section used in this analysis is $6.18 \times 10^4$ pb ($1.979 \times 10^4$ pb) for $W^\pm+$jets ($Z+$jets)~\cite{ATLAS-CONF-2015-041} at NNLO. Finally, the cross section for QCD multijets backgrounds are taken from the Madgraph. For the signal, we have used the LO cross section calculated from the Madgraph to give a conservative collider reach.

The signal topology as in eq.~\ref{di-top_signal} is shown in right panel of fig.~\ref{fig:di-toptopology} where each $U_1$ vector LQ decays to the top quark and anti-neutrino. Subsequently, the top quark decays hadronically and the decay products are collimated because they are produced from a highly boosted top quark. The deep green blobs are diagrammatic representation of the large radius jets denoted as $t_j$ ($\bar{t}_j$) from the top (anti-top). 

Several kinematic variables have been used in our analysis which utilize the available kinematic information to maximize the significance. They are: missing transverse energy ($\slashed{E}_T$), transverse mass variable $M_{T2}$~\cite{Lester:1999tx, Barr:2003rg, Meade:2006dw, Lester:2007fq, Cho:2007qv, Cho:2007dh, Barr:2007hy, Gripaios:2007is, Nojiri:2008hy, Konar:2009qr}, $\sqrt{\hat{s}}_{min}$~\cite{Konar:2008ei, Konar:2010ma, Papaefstathiou:2009hp, Papaefstathiou:2010ru, Swain:2014dha, Swain:2015qba}, razor variables~\cite{Rogan:2010kb, Buckley:2013kua, Chatrchyan:2011ek, Khachatryan:2015pwa}, $H_T$~\cite{Tovey:2000wk} and $M_{eff}$~\cite{Hinchliffe:1996iu} which we discuss briefly in what follows.

\begin{figure}[t]
	\centering
        \includegraphics[keepaspectratio=true,scale=0.58]{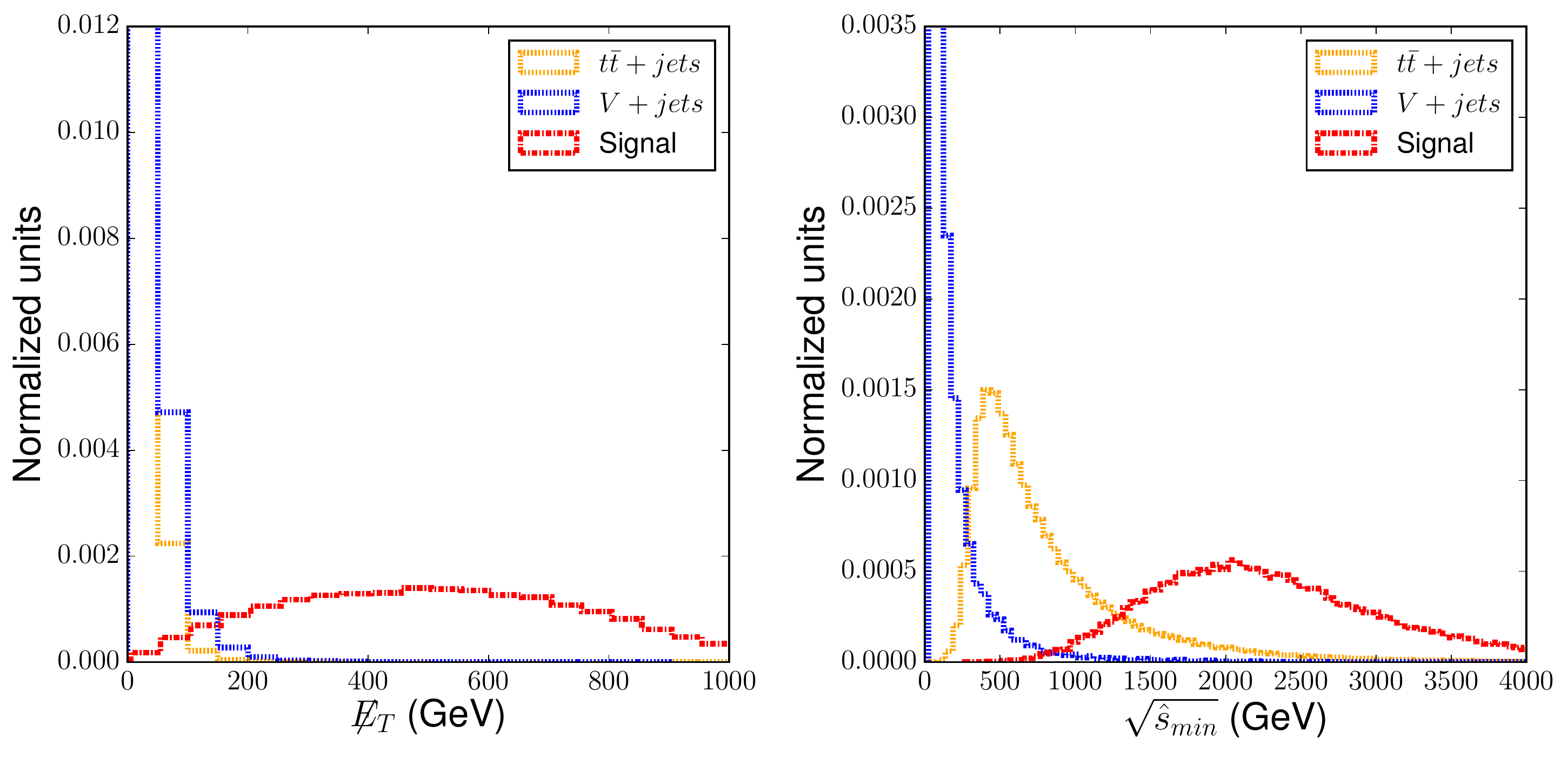}
	\caption{The variable missing transverse energy (left panel) and $\sqrt{\hat{s}}_{min}$ (right panel) are displayed here. The signal, in red colour, here corresponds to the pair production of the $U_1$ vector LQ with mass 1 TeV. The dominant backgrounds, $t\bar{t} + jets$,  $V + jets$ with $V = W^{\pm}/Z$ are displayed in orange and blue respectively.}
	\label{fig:MET_shatmin}
\end{figure}

The missing transverse energy, $\slashed{E}_T$, is the momentum imbalance in the transverse direction. It is expected to have a significant value subject to the presence of invisible particles in the final state. Otherwise, it attains a comparatively smaller non-zero value owing to mismeasurement. Since the signal considered here has neutrinos in the final state, they generate a significant amount of missing energy. The missing energy corresponding to the background events, however, is mostly due to mismeasurement except for some small fraction of events where neutrinos contribute. Fig.~\ref{fig:MET_shatmin}, (left panel) shows the distribution of missing transverse energy for signal and backgrounds where the red colour corresponds to signal and only the dominant backgrounds are displayed. The mass of the $U_1$ vector LQ in this representative plot is taken to be 1 TeV. One can immediately verify that $\slashed{E}_T$ for the signal peaks at a value that is much higher compared to the backgrounds where it peaks at values close to zero. Hence, $\slashed{E}_T$ is an important variable suitable for handling the large SM backgrounds.

The mass bound variable $\sqrt{\hat{s}}_{min}$~\cite{Konar:2008ei, Konar:2010ma, Papaefstathiou:2009hp, Papaefstathiou:2010ru, Swain:2014dha, Swain:2015qba} has been proposed to measure the mass scale associated with NP. This is a global and inclusive variable which can be applied for any event topology without caring about the number of parent and the number of invisible particles involved in the topology. When there are invisible particles present in the final state, it is very challenging to get the information of the partonic center of momentum (CM) energy, $\sqrt{\hat{s}}$, which is nothing but the mass of the heavy resonance for singly production or the threshold of the pair production. $\sqrt{\hat{s}}_{min}$ is an interesting way out where the peak (end-point) of the distribution is nicely correlated with  the pair production (singly produced heavy resonance). For a given event, it is defined as the minimum partonic CM energy that is required to produce the given final state particles and the measured missing transverse energy. Mathematically, 
\begin{eqnarray}\label{shatMin}
\sqrt{\hat{s}_{\min}(m_{inv})} = \sqrt{(E^{vis})^2 - (P_z^{vis})^2} + \sqrt{\mptvec^2 + m_{inv}^2}, 
\end{eqnarray}
where $m_{inv}$ is sum of the invisible particle masses while $E^{vis} = \sum_j e^{vis}_j$ is the total visible energy and $P_z^{vis} = \sum_j p^z_{j}$ stands for total longitudinal component of the visible momenta of the reconstructed objects. The above expression for $\sqrt{\hat{s}}_{min}$ is obtained after minimizing the partonic Mandelstam variable $\hat{s}$ with respect to the invisible momenta subject to the missing transverse momentum constraints. This is a function extremization problem which can be done analytically as well as numerically, we have checked it numerically using Mathematica and found that the result exactly matches with the one mentioned in eq.~\ref{shatMin}. It turns out that the first term depends on the visible decay products but the second term depends on the missing transverse energy $\mET$ and the sum of the invisible particle mass, $m_{inv}$ after the minimization. Hence, the variable $\hat{s}_{min}$ is a function of the sum of the masses of the invisible particles in the final state. However, fortunately, in our case the invisible particles are only neutrinos which makes $\hat{s}_{min}$ independent of $m_{inv}$ and it only depends on the visible momenta and missing transverse energy. Fig.~\ref{fig:MET_shatmin}, (right panel) shows the distribution of $\sqrt{\hat{s}}_{min}$. Similar to the earlier case, the distribution in red corresponds to the signal for a $U_1$ vector LQ of mass 1 TeV. By construction, $\sqrt{\hat{s}}_{min}$ peaks at the threshold for the pair production. Considering the pair production of  $U_1$ as our signal, the peak at 2 TeV hence matches well with the theoretical expectation for the variable. Since the threshold for the backgrounds are much smaller compared to the signal, this variable is also a smart choice as far as reducing the SM backgrounds is concerned.   

\begin{figure}[t]
	\centering
        \includegraphics[keepaspectratio=true,scale=0.58]{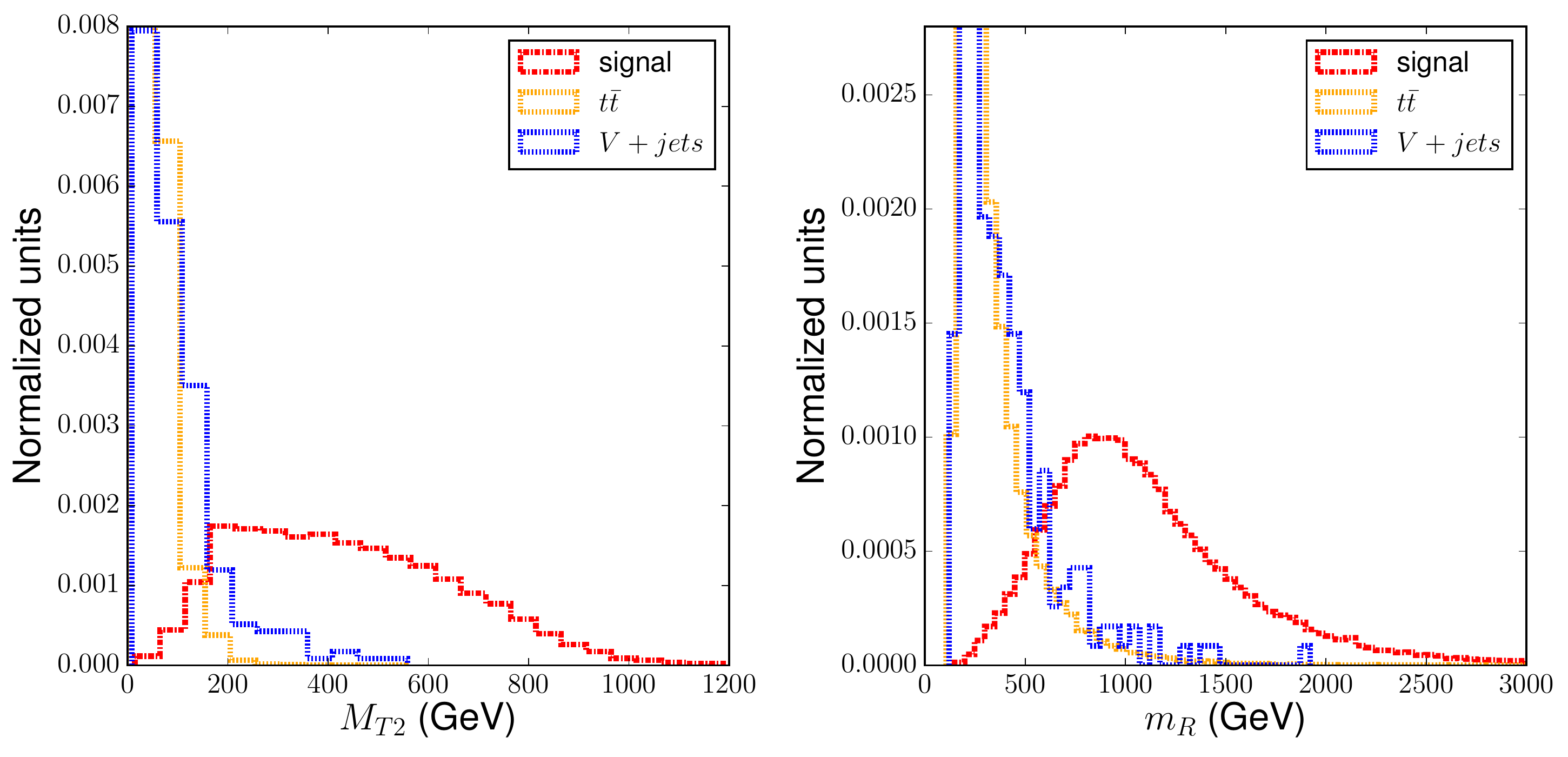}
	\caption{The variables $M_{T2}$ (left panel) and $m_R$ (right panel) are displayed where the red coloured histogram corresponds to the signal for a $U_1$ vector LQ of mass 1 TeV. As discussed in the text, the endpoint corresponding to $M_{T2}$ falls at $M_{U_1}=$1 TeV and $m_R$ peaks at $M_{U_1}$.}
	\label{fig:Razor_MT2}
\end{figure}

The $(1+2)$ dimensional transverse mass variable, $M_{T2}$~\cite{Lester:1999tx, Barr:2003rg, Meade:2006dw, Lester:2007fq, Cho:2007qv, Cho:2007dh, Barr:2007hy, Gripaios:2007is, Nojiri:2008hy, Konar:2009qr}, plays a pivotal role in reducing the background events. As a result, the signal significance is satisfactorily enhanced even though this variable was initially defined for the mass measurement of new particles both in long and short decay chains. The kinematic variable $M_{T2}$ is defined as the maximum transverse mass between the two parents satisfying the missing transverse momenta ($\mptvec$) constraints and then minimizing over the momenta of the invisible particles (e.g., neutrino).
\begin{equation}\label{mt2}
M_{T2} (m_{\nu})     \equiv     \min_{\substack{\vec{q}_{iT} \\ \sum \vec{q}_{iT} = \vec{\slashed{P}}_T }} 
\left[  \max_{i =1, 2}      \{  M_T^{(i)}  ({p}_{iT}, {q}_{iT}, m_{vis(i)}; m_{\nu})  \}  \right],
\end{equation}
where the $M_T^{(i)}$ for each decay chain are,
\begin{eqnarray}
&&(M_T^{(i)})^2 = m_{vis(i)}^2 + m_{\nu}^2 + 2(E_T^{vis(i)}E_T^{\nu(i)} - \vec{p}_{iT}.\vec{q}_{iT})\;,\\  \label{mt}
&&E_T^{vis(i)} = \sqrt{m_{vis(i)}^2 + p_{iT}^2}, \, \,\,\,E_T^{\nu(i)} = \sqrt{m_{\nu}^2 + q_{iT}^2}  \, \, .
\end{eqnarray}
In the above $\vec{p}_{iT}$ and $\vec{q}_{iT}$,  $E_T^{vis(i)}$ and $E_T^{\nu(i)}$ are the transverse momentum, transverse energy of the large radius jet from top (anti-top) and neutrino respectively. Note that the visible quantities in each event for the signal, are the two hardest $P_T$ large radius jets (hardest $P_T$ jets). The remaining reconstructed quantities are assumed to be soft and do not change the $M_{T2}$ distribution significantly.
Note that the minimization, in the definition of $M_{T2}$, acts over all the partitions of missing transverse momenta constraints. The maximization, on the other hand, is done between the two transverse masses for each partition. This ensures that the resulting $M_{T2}$ gets closer to the $U_1$ mass, $M_{U_1}$. By construction, $M_{T2}\leq M_{U_1}$ where the equality holds when the top (anti-top) quark and the anti-neutrino (neutrino) are produced with equal rapidity. Hence,  for the correct input mass of the invisible daughter particle, the endpoint for $M_{T2}$ is at the mass of the $U_1$ vector LQ $M_{U_1}$. The neutrino mass being very small, we assume it to be zero for the $M_{T2}$ calculation. In fig.~\ref{fig:Razor_MT2}, (left panel) the $M_{T2}$ distribution is displayed where the red colour corresponds to the signal. Since the mass of the $U_1$ vector LQ is taken to be 1 TeV, the end-point of the distribution, as expected, is at the same value albeit with very small number of events. Most of the backgrounds fall sharply at around 200 GeV which makes this mass bound variable extremely important in maximizing the signal to background ratio.

\begin{figure}[ht]
	\centering
\includegraphics[keepaspectratio=true,scale=0.48]{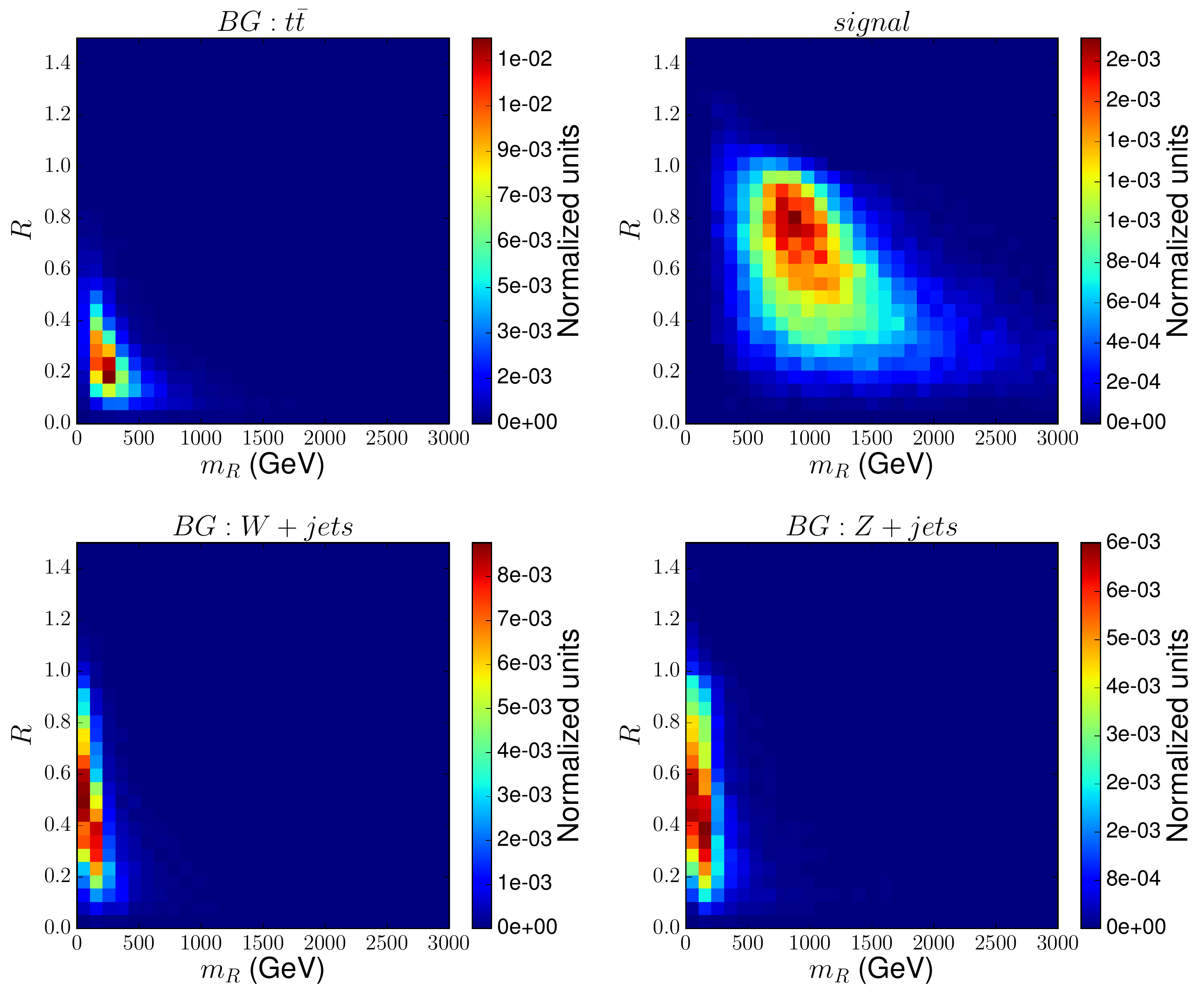}
	\caption{The dimensionless ratio of razor variables, $R$, in the y-axis and $m_R$ in the x-axis with normalized events in the coloured bar represented for the signal and for the dominant backgrounds. By construction, as discussed in the text, the variable $R$ peaks at higher values for the signal and for the backgrounds however, it peaks near zero. The variables $R$ and $m_R$ are very effective in distinguishing the signal from the background events.}
	\label{fig:Razor_R_MR}
\end{figure}

The razor variable~\cite{Rogan:2010kb, Buckley:2013kua, Chatrchyan:2011ek, Khachatryan:2015pwa} is another interesting observable well known for handling SM backgrounds with di-jet\footnote{In this analysis we have selected events with at least two large radius jets to calculate the razor variables for both the signal and backgrounds. For more than two large radius jets we select the two which are the hardest.} and missing transverse energy produced from the pair production of heavy resonance. Assuming the heavy resonances are produced at the threshold, which is true for many BSM scenarios except the cases when the resonance is not so heavy, one calculates the two following  mass variables:
\begin{eqnarray}\label{MRandMTR}
 && m_R = \sqrt{(|\vec{p}_{j1}| + |\vec{p}_{j2}|)^2 - (p_{j1}^{z} + p_{j1}^{z})^2},\label{MR}\\
 && m_{TR} = \sqrt{\frac{1}{2}[|\mptvec|(p_{T}^{j1} + p_{T}^{j2}) - \mptvec . (\vec{p}_{T}^{j1} + \vec{p}_{T}^{j2})]}\label{MTR}.
\end{eqnarray}
In the razor frame, the longitudinal component of the momentum of the two visible decay products are equal and opposite. With this assumption, the variable $m_R$ will display a peak at the mass of the $U_1$ vector LQ, $M_{U_1}$ (where the neutrino mass is assumed to be zero). The transverse mass, $m_{TR}$, contains the information of the missing transverse energy due to neutrinos for signal events. The missing transverse energy for most of the background events is due to mismeasurement. Although, there are some background events which contain neutrino(s) in the final state, the number of events with such a final state is very small statistically and does not contribute much. Intuitively for the signal events transverse mass ($m_{TR}$) is smaller or equal to the $U_1$ vector LQ mass ($M_{U_1}$), i.e.,  $m_{TR}\leq M_{U_1}$, but such a relation will not hold for the background events. In order to better discriminate the signal and background events a dimensionless ratio is defined as follows,
\begin{equation}
R \equiv \frac{m_{TR}}{m_R}.
\end{equation}
While $R$ for backgrounds will peak at zero, for the signal it will peak at higher values giving a better discrimination between the two. The variable $m_R$ is represented in fig.~\ref{fig:Razor_MT2} (right panel). It is immediately evident that it peaks at the mass of the $U_1$ vector LQ for the signal events. The corresponding peak for backgrounds is at comparatively smaller values. The dimensionless ratio $R$ is displayed in fig.~\ref{fig:Razor_R_MR} which represents a 2 dimensional histogram where the colour axis represents the normalized events. The variable $m_R$ along with the dimensionless ratio $R$ is appropriate for handling the background events efficiently. Since the signal peaks at higher values of $R$ and $m_R$ in the $R - m_R$ plane compared to the backgrounds, a moderate cut on both $R$ and $m_R$ would be sufficient in order to minimize the backgrounds.

In addition to the above kinematic observables, there are some global and inclusive variables like the previously discussed $\sqrt{\hat{s}}_{min}$ but independent of any parameter related to the invisible particle. Out of many similar observables from the literature which are extensively utilized in experimental analyses we have considered two of them, $H_T$~\cite{Tovey:2000wk} and $M_{eff}$~\cite{Hinchliffe:1996iu}. The variable $H_T$ is defined as the scalar sum of the transverse momenta of the visible decay products in the final state. In this analysis there only jets which are visible, so we have constructed $H_T$ by doing the scalar sum of the transverse momentum of the jets. This is an inclusive variable which means it is unaffected by the combinatorial ambiguity of jets. It was proposed to give the information about the mass scale of new physics where there are large number of jets present in the final state. The variable $M_{eff}$ is defined as the scalar sum of transverse momenta of jets and missing transverse energy, $M_{eff} = \sum_i P_T^{i}(j) + \mET$. This variable also measures the mass scale of NP when there are invisible particles associated with it.

\begin{figure}[t]
	\centering
        \includegraphics[width=80mm,height=59.0mm]{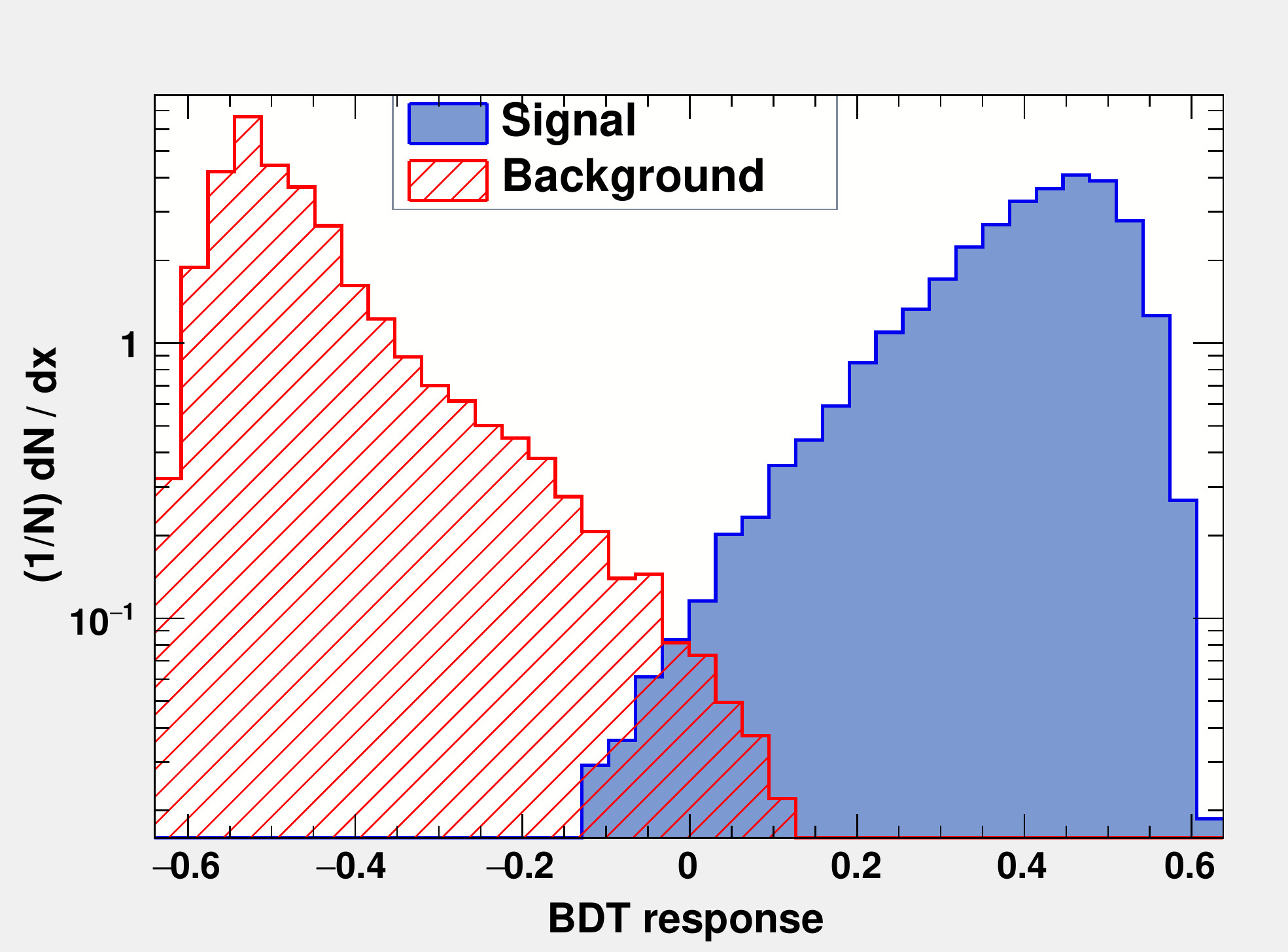}
	\includegraphics[width=80mm,height=59.5mm]{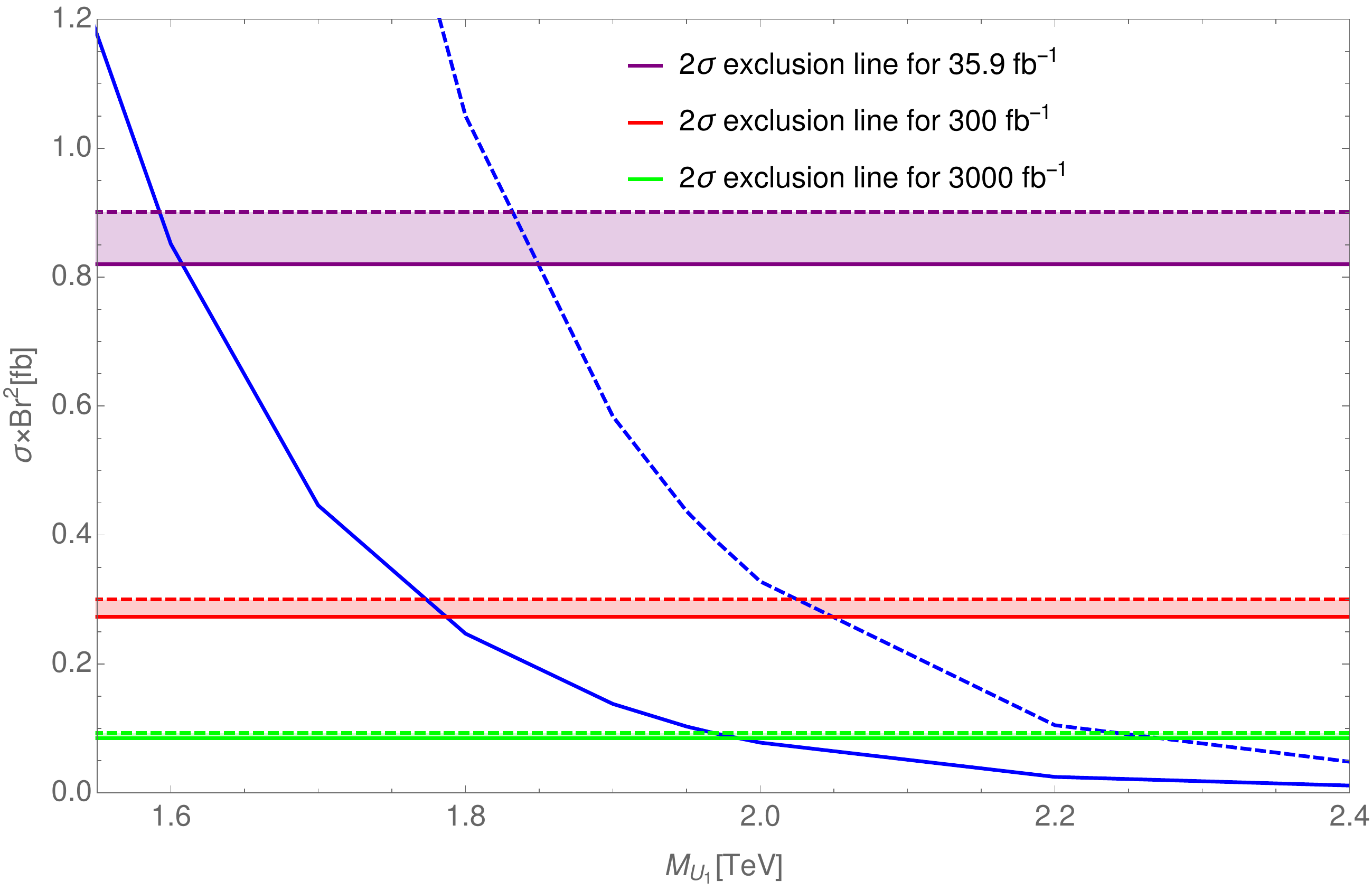}
	\caption{ The figure (left panel) displays the BDT response for the considered signal (solid blue region) and background (red dashed region). As evident from the figure, BDT disentangles the signal from background quite efficiently. The exclusion limit (right panel) at $95\%$ C.L. for the 13 TeV CM energy is presented for the signal. The exclusion limits are calculated for luminosities 35.9 $\rm{fb}^{-1}$ (purple band), 300 $\rm{fb}^{-1}$ (red band) and 3000 $\rm{fb}^{-1}$ (green band) respectively. The corresponding bands represent 20$\%$ systematic uncertainty in the background events. The dashed and solid blue lines represent the theoretical cross section times branching ratio square that are computed for 100\% and 50\% branching ratio of the vector LQ ($U_1 \to t ~ \bar{\nu}$) respectively.}
	\label{significance}
\end{figure}

We disentangle the signal and background by multivariate technique using toolkit for Multivariate data analysis (TMVA) with {\tt Root}, particularly Boosted Decision Tree (BDT). We utilize the above discussed (efficient) variables along with some other interesting exclusive variable to maximize the signal significance. The following feature vectors are employed as input to BDT: $\mET$, $H_T$,~$M_{eff}$,~$\sqrt{\hat{s}}_{min},~M_{T2},~M_R,~ R, ~ N_b, ~ N_{fatJet} ~\text{and}~  N_j$. Where, the variables $N_b$ and $N_j$ are defined as the number of b$-$jets and number of jets respectively present in each event. The BDT response is displayed in fig.~\ref{significance} (left panel) where the red histogram corresponds to background events and the blue filled histogram is for the signal. Evidently, the signal and background distributions are well separated which in turn maximizes the LHC reach for the $U_1$ vector LQ.	
%
The signal and background events are calculated after putting an optimal BDT cut value, we have then enumerated the statistical significance of the aforementioned signal using the following formula:
\begin{eqnarray}
{\cal S} = \frac{N_s}{\sqrt{N_s+ N_b(1 + \epsilon_b)}}\;\;.
\label{eq:signi}
\end{eqnarray}
Here, $N_s(N_b)$ denotes the number of signal (background) events after implementing the BDT cut value for a specific luminosity and $\epsilon_b$ is systematic uncertainty~\cite{sig:un}. The TMVA  analysis gives background and signal cut efficiency as output for any given cut value of the method (here we have used BDT). It is then quite straight forward to calculate the number of signal ($N_s$) and background events ($N_b$) using corresponding cut efficiency and then the significance from eq.~\ref{eq:signi}. Using eq.~\ref{eq:signi}, we obtained a marginally improved expected limit of 1600 GeV for 13 TeV LHC with 35.9 $\rm{fb}^{-1}$ integrated luminosity ($L_{int}$) compared to the CMS analysis \cite{Sirunyan:2018kzh} where the observed (expected) exclusion limit at $95\%$ C.L. was 1530 (1460) GeV. We find that the vector LQ can be excluded up to 1780 (1980) GeV  at $95\%$ C.L. for 13 TeV LHC with 300 (3000) $\rm{fb}^{-1}$ integrated luminosity. The above limits were computed for the vector LQ decaying to di-top plus missing energy channel with a $50\%$ branching ratio. The LHC reach at $95\%$ C.L. with $100\%$ branching ratio are 1830, 2020 and 2230 GeV for 35.9, 300 and 3000 $\rm{fb}^{-1}$ integrated luminosity respectively.
In fig.~\ref{significance} (right panel) we depict the exclusion limit at $95\%$ C.L. for $\sqrt{s} = 13$ TeV. The red and green bands denote the exclusion limits for 300 and 3000 $\rm{fb}^{-1}$ integrated luminosity respectively. The blue solid (dashed) line denotes the effective theoretical production cross-section for $50\%$ ($100\%$) branching ratio, at the leading order, with the variation of the mass of $U_1$ vector LQ. The LHC reach for scalar LQ (see lower panel of Figure no.~3 of ref.~\cite{Sirunyan:2018kzh}) decaying to the same channel, di-top plus missing energy, with $100\%$ branching ratio at 13 TeV LHC is 1020 GeV for 35.9 $\rm{fb}^{-1}$ integrated luminosity. Evidently, the stronger LHC reach for the $U_1$ vector LQ is mainly because of the higher production cross section compared to scalar leptoquark. Although, the choice $\kappa = 0$ and with $50\%$ branching ratio will reduce the LHC limit for the vector LQ because of slightly smaller cross section but still larger than scalar LQ pair production. Hence, the vector LQ LHC reach may always be stronger than that of the scalar LQ for pair production in di-top plus missing energy channel. In addition, we have also taken 20$\%$ systematic uncertainty in the background estimation and calculated the significance. After the inclusion of systematic uncertainty, the significance is reduced slightly which can be seen from the bands (purple, red and green for 35.9, 300 and 3000 $\rm{fb}^{-1}$ integrated luminosity respectively) in right panel of fig.~\ref{significance}. The mass of $U_1$ vector LQ can now be excluded up to 1585 GeV at $95\%$ C.L. for 13 TeV LHC with 35.9 $\rm{fb}^{-1}$ integrated luminosity when the LQ decays to di-top plus missing energy channel with 50$\%$ branching ratio. This limit can reach up to 1770 and 1975 GeV for 300 and 3000 $\rm{fb}^{-1}$ integrated luminosity respectively.

\section{Constraints from $b\to c\tau\nu_l$ observables}\label{flav}
Flavour physics has been instrumental in the search for NP which has been the main interest of the current phenomenological community for the last decade. The $\mathcal{R}_{D^{(*)}}$ and $\mathcal{R}_{K^{(*)}}$ ratios with deviations of about $3.08\sigma$ and $2.6\sigma$ from their SM values respectively, along with other observables, have been much discussed as probes for such LFUV NP. However, since the $U_1$ vector LQ has been assumed to couple to third generation leptons, it will not contribute to the $\mathcal{R}_{K^{(*)}}$ anomalies. Hence, we carry out an analysis on the scope of the $U_1$ LQ in explaining the current $\mathcal{R}_{D^{(*)}}$ anomaly. Furthermore, in the case of $\mathcal{R}_{D^{(*)}}$ observable, due to the above mentioned assumption the NP couple only to the numerator (i.e., to the third generation leptons) while the denominator stays SM like. This approach has been widely followed in the community and is also the one that we follow in the flavor section of our analysis. We consider the data due to the different collaborations as separate data points instead of using the global average for these results. This results in an increase of statistics and degrees of freedom. It also allows us to take care of correlations unaccounted for in the global average, such as that between the 2016-2017 Belle measurement and the $D^*$ polarization $P_{\tau}(D^*)$. We perform a fit to a total of 11 observables, including the fraction of the longitudinal polarization of $D^{(*)}$ ($F_L(D^*)$), the $\tau$ polarization in a $B\to D^*$ decay ($P_{\tau}(D^*)$) and the most recent Belle $\mathcal{R}_{D^{(*)}}$ measurements\footnote{We refrain from using the $R_{J/\psi}$ measurements since the corresponding SM estimates are far from accurate due to the absence of a reliable form factor parameterization for these decays.}. The observables we use in our fit are listed in table~\ref{tab:obs}. The corresponding SM estimates are also mentioned on the topmost row, whereas the estimates for the $U_1$ vector LQ model are provided in the bottom row of the same table. 
	\begin{table}[ht]
		\begin{flushleft}
		     \tiny
				\begin{tabular}{cccccc}
					& $\mathcal{R}(D)$  & $\mathcal{R}(D^*)$  &	Correlation	& $P_{\tau}(D^*)$ &$F_L(D^*)$\\
					\hline
					SM			& $0.305(3)$		& $0.260(9)$	 	                &				          &  $-0.486(29)$& $0.459(11)$\\
					\hline
 					Babar~\cite{Lees:2012xj,Lees:2013uzd}   	 & $0.440(58)_{st.}(42)_{sy.} $ & $0.332(24)_{st.}(18)_{sy.}$           & $-0.31$ &              &\\
 					Belle (2015)~\cite{Huschle:2015rga}     & $0.375(64)_{st.}(26)_{sy.}$ 	& $0.293(38)_{st.}(15)_{sy.}$           & $-0.50$ &              &\\
 					Belle (2016)~\cite{Hirose:2016wfn,Hirose:2017dxl}     & - 		        	& $0.270(35)_{st.}~^{+ 0.028}_{-0.025}$ & 0.33                                    & $ -0.38(51)_{st.}~^{+0.21}_{-0.16}$ &\\
  					Belle (2019)~\cite{Abdesselam:2019dgh}     & $0.307(37)_{st.}(16)_{sy.}$  & $0.283(18)_{st.}(14)_{sy.}$           & $-0.51$                                 &                                     &\\
 					LHCb (2015)~\cite{Aaij:2015yra}      & - 				& $0.336(27)_{st.}(30)_{sy.}$  & & &\\
 					LHCb (2017)~\cite{Aaij:2017deq,Aaij:2017uff}      & - 				& $0.280(18)_{st.}(29)_{sy.}$  & & &\\
 					\hline
 					World Avg.~\cite{Adamczyk:2019wyt}       & $0.340(27)_{st.}(13)_{sy.}$	& $0.295(11)_{st.}(8)_{sy.}$ & $-0.38$  & $ -0.38(51)_{st.}~^{+0.21}_{-0.16}$&$0.60(08)_{st.}(035)_{sy.}$\\\hline
					$U_1$ vector LQ Values. & $0.347(51)$	& $0.297(20)$ &   &$-0.486(25)$ &$0.459(12)$\\\hline
				\end{tabular}
 				\caption{Present status (both theoretical and experimental) of $\mathcal{R}(D)$, $\mathcal{R}(D^*)$, $P_{\tau}(D^*)$ and $F_L(D^*)$. First uncertainty is statistical and the second one is systematic. The first row lists the SM calculation obtained in this paper, while the last row includes the estimates for the $U_1$ vector LQ model after the fit.} 
				\label{tab:obs}
		\end{flushleft}
	\end{table}

The effective Lagrangian for a $b\to c\tau\nu_l$ decay with all possible vector and scalar Wilson coefficients (WCs) with left-handed neutrinos can be written as~\cite{Sakaki:2013bfa}:
\begin{equation}\label{Leff}
   H_{eff} = {4G_F \over \sqrt2} V_{cb}\left[ (\delta_{l\tau} + C_{V_1}^l)\mathcal{O}_{V_1}^l + C_{V_2}^l\mathcal{O}_{V_2}^l + C_{S_1}^l\mathcal{O}_{S_1}^l + C_{S_2}^l\mathcal{O}_{S_2}^l + C_T^l\mathcal{O}_T^l \right] \,,
\end{equation}
where $G_F$ is the Fermi constant for weak interactions, $V_{cb}$ is the relevant Cabibbo-Kobayashi-Maskawa (CKM) element for $b\to c$ quark transitions. The $U_1$ vector LQ does not contribute to all of the above mentioned WCs, but only to $C^l_{V_1}$ and $C^l_{S_1}$. In accordance with~\cite{Sakaki:2013bfa} which provides the complete list of WCs relevant for LQ models and contributing to $b\to c\tau\nu$ decays and the corresponding operator basis, the $b\to c\tau\nu_l$ WCs relevant for the $U_1$ vector LQ can be written as\footnote{Our notation for the LQ couplings are slightly different from the one generally used in the literature. For example, $h^{cl}_{1L}$ is generally written as $h^{2l}_{1L}$ where the $1$ in the superscript represents quarks from the second generation. However, for $l=\mu$, the second term in the numerator of $C^l_{V_1}$ is generally written as $h^{k2}_{1L}$ but this time the number $2$ represents first generation leptons. To avoid such confusions, we label the couplings using letters corresponding to the various quark and lepton generations where $l={e,\mu,\tau}$}:
\begin{eqnarray}\label{LQ_WC}
C^l_{V_1}&=&\frac{1}{2\sqrt{2}G_FV_{cb}}\sum_{k=1}^3V_{k3}\frac{h^{cl}_{1L}h^{kl}_{1L}}{M^2_{U^{2/3}_1}},\nonumber\\
C^l_{S_1}&=&-\frac{1}{2\sqrt{2}G_FV_{cb}}\sum_{k=1}^3V_{k3}\frac{2h^{cl}_{1L}h^{kl}_{1R}}{M^2_{U^{2/3}_1}},
\end{eqnarray}
where $l={e,\mu,\tau}$ in general, $V_{k3}$ denotes the CKM elements and the upper index of the LQ denotes its electric charge. We discard the contribution from the Cabibbo suppressed terms and keep the leading terms proportional to $V_{33}=V_{tb}$. 

The theoretical expressions for the observables $\mathcal{R}(D^{(*)})$ and the $\tau$ polarization $P_{\tau}(D^*)$ have been obtained from ref.~\cite{Sakaki:2013bfa}, and for the fraction of the longitudinal polarization of the $D^*$ meson $F_L(D^*)$ from ref.~\cite{Bhattacharya:2018kig}. These expressions depend on the partial $q^2$ dependent decay widths for the $B\to D^{(*)}\tau\nu_l$ decays, where $q^2$ is the di-lepton invariant mass. These widths have also been taken from ref.~\cite{Sakaki:2013bfa} and, for the sake of completion, have been provided in Appendix~\ref{obs_def}. In order to maintain parity with the collider section, we assume the WCs to be real. We also assume the $U_1$ vector LQ to couple to only the third generation leptons, and hence $l=\tau$ in the following.

\begin{figure}[ht]
\centering
\subfloat[]{\label{fit}\includegraphics[width=0.4\linewidth]{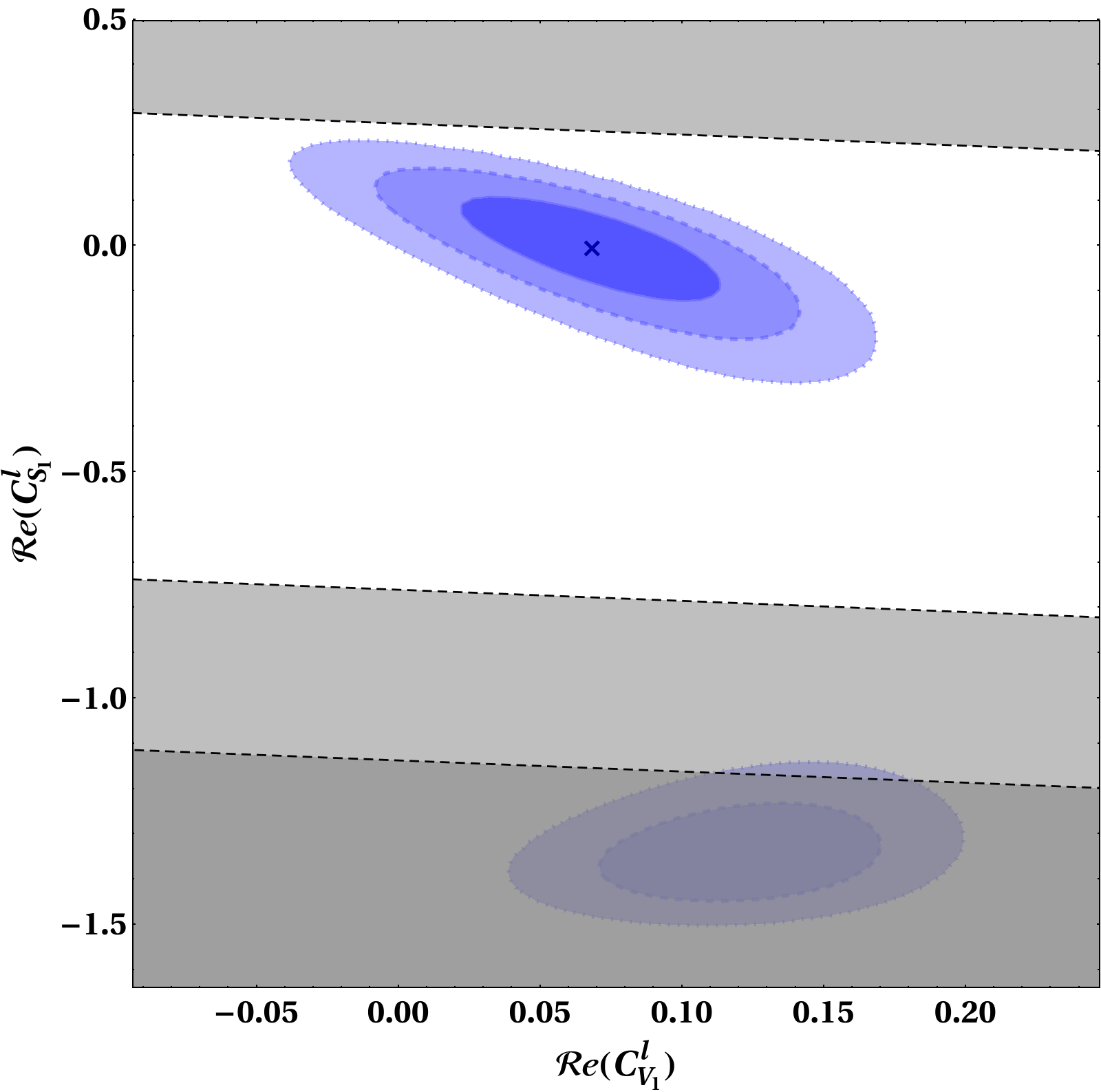}}\hfill
\subfloat[]{\label{model}\includegraphics[width=0.5\linewidth]{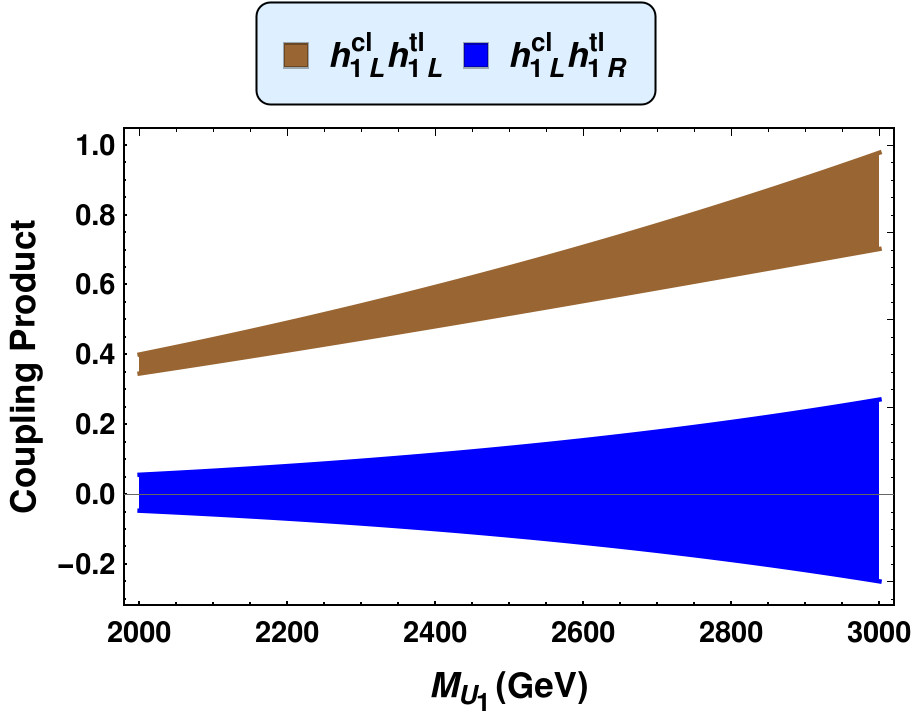}}\\
\caption{Our fit results displayed in the $\mathcal{R}e(C^\ell_{S_1})$ vs $\mathcal{R}e(C^\ell_{V_1})$ plane (fig.~\ref{fit}). One can see that there are two best fit regions. One of them is discarded by both the relaxed ${\rm Br}(B_c\leq30\%)$~\cite{Alonso:2016oyd} (shown in dark grey) and the aggressive ${\rm Br}(B_c\leq10\%)$~\cite{Akeroyd:2017mhr} (displayed in light grey) bounds. The best fit point is marked by a cross and the $1$, $2$ and $3\sigma$ contours corresponding to the best fit points are shown. The $1\sigma$ contour translates to the $1\sigma$ bands for the coupling products $h^{cl}_{1L}h^{kl}_{1L}$ and $h^{cl}_{1L}h^{kl}_{1R}$ in the model parameter space, displayed in brown and blue respectively in fig.~\ref{model}. Here $l=\tau$. The range for the mass of the $U_1$ vector LQ (the x-axis in fig.~\ref{model}) is in accordance with the findings from the collider section of this work (see the last line of the second last paragraph under Introduction (sec.\;\ref{intro})).}
\end{figure}

	\begin{table}[ht]
		\begin{center}
				\begin{tabular}{cccc}
					& Parameters  & Fit Values&\\
					\hline
					& $C^l_{V_1}$   &  $0.068\pm0.03$&        \\
                                        & $C^l_{S_1}$   &  $-0.0019^{+0.0806}_{-0.0844}$&       \\\hline
					\end{tabular}
 				\caption{The fit results for the NP parameters along with their $1\sigma$ errors. Here $l=\tau$.} 
				\label{tab:fit}
		\end{center}
	\end{table}
The results from our fits are displayed in fig.~\ref{fit}. One can see that there are two best-fit regions. However, one of them is discarded by the constraints due to the not yet measured branching ratio (Br) for the $B_c\to\tau\nu$ mode. It has been well known in the literature that pure scalar NP as an explanation for the $b\to c\tau\nu$ anomalies is highly constrained due to even a relaxed limit of $B_c\to\tau\nu\leq30\%$~\cite{Alonso:2016oyd}. In our current analysis, we use this relaxed as well as a more aggressive bound of ${\rm Br}(B_c\to\tau\nu)\leq10\%$ obtained from the LEP data taken at the Z-peak~\cite{Akeroyd:2017mhr}. We see that both these constraints amount to one of the two best-fit regions being discarded. The best-fit points that survive this constraint along with the corresponding $1\sigma$ errors are shown in table~\ref{tab:fit}\footnote{It should be mentioned here that a recent work~\cite{Blanke:2018yud,Blanke:2019qrx} disagrees with the $10\%$ bound, claiming it to be too aggressive. They also argue that a more relaxed limit of $60\%$ for the same branching ratio is allowed. However, in order to keep our analysis as aggressive as possible, we do not use the $60\%$ limit. Using this limit will result in the discarded region being allowed, and will hence result in a more loosely constrained model parameter space. The values of the observables in the $U_1$ vector LQ model will, however, remain unchanged.}. This $1\sigma$ region translates into the $1\sigma$ bands in the model-parameter space for the $U_1$ vector LQ depicted in fig.~\ref{model}.
 
From fig.~\ref{model} as well as table~\ref{tab:fit}, it is clear that $C^{l}_{S_1}$ is consistent with its SM value of $0$. This is primarily due to the fact that the inclusion of the most recent Belle (2019) results shifts the global average closer to the SM estimates for the $\mathcal{R}_{D^{(*)}}$ ratios, bringing the global deviation between the SM and the experimental results down from $\sim 4$ to $\sim 3\sigma$. The non-zero $C_{V_1}$ is enough to explain such deviations in the $\mathcal{R}_{D^{(*)}}$ ratios. However, a glimpse at table~\ref{tab:obs} shows that the NP estimates corresponding to the $P_{\tau}(D^*)$ and the $F_L(D^*)$ observables for the $U_1$ vector LQ model are consistent with the corresponding SM estimates. While the NP estimates for $P_{\tau}(D^*)$ are also consistent with the corresponding experimental value within $1\sigma$ (primarily due to the large statistical error associated with its experimental value), the $F_L(D^*)$ NP estimate is off from its experimental value by almost $1.6\sigma$.
\section{Conclusion}\label{concl}
In this article, we consider a particular type of leptoquark scenario that contains
the $U_1$ leptoquark which conserves baryon and lepton numbers. 
This leptoquark mediates both charged as well as neutral current processes
involved in the $B$-physics anomalies at tree level. In this work we have performed a comprehensive collider analysis of the $U_1$ 
vector leptoquark via multijet plus missing transverse energy final states. Further, in our current article we have assumed that the $U_1$ vector leptoquark couples with only third generation of leptons but with second and third generations of quarks, it will not contribute to the $b\to sll$ anomalies. We have hence studied the effect of such leptoquark in $b\to c$ observables only.

In addition, all the couplings share the same values. 
We have constructed several non-trivial kinematic variables which help us to 
reduce the SM background with respect to the signal in our collider analysis. We then implemented the interesting kinematic variables in a multivariate analysis, BDT, to maximize the LHC reach for the $U_1$ vector leptoquark. From our study, we have derived exclusion mass limits for the $U_1$ leptoquark at 95\% C.L. corresponding to the 13 TeV LHC run with two benchmark values for integrated luminosities. For example,  we can exclude up to 2020 (1780) GeV for an integrated luminosity of 300 $\rm{fb}^{-1}$, 2230 (1980) GeV for 3000 $\rm{fb}^{-1}$ when the vector leptoquark decays with $100\%$ ($50\%$) branching ratio to the di-top plus missing energy channel. 

We also analyze the scope of the $U_1$ leptoquark in explaining the $b\to c\tau\nu$ anomalies. We exclude the $R(J/\psi)$ data since the corresponding theoretical estimates are inaccurate, primarily due to lack of a proper form factor parameterization. We find that the $U_1$ leptoquark can potentially explain these results even after the more aggressive constraint of ${\rm Br}(B_c\leq\tau\nu)$ has been applied. 

\section{Acknowledgement}
NG would like to acknowledge the Council of Scientific and Industrial Research (CSIR), Government of India for financial support. The work of AKS was supported by the Department of Science and Technology, Government of India under the fellowship reference number PDF/2017/002935 (SERB NPDF).

\begin{appendices}
\renewcommand{\thesection}{\Alph{section}}
\renewcommand{\theequation}{\thesection-\arabic{equation}} 
\setcounter{equation}{0}
\section{$\mathbf{B\to D^{(*)}}$ observables}\label{obs_def}
The differential decay rate for $B\to D^{(*)}\tau\nu_l$ decays can be written as:
\begin{equation}
   \begin{split}
      {d\Gamma(\Bbar \to D \tau\nubar_l) \over dq^2} =& {G_F^2 |V_{cb}|^2 \over 192\pi^3 m_B^3} q^2 \sqrt{\lambda_D(q^2)} \left( 1 - {m_\tau^2 \over q^2} \right)^2 \times\biggl\{ \biggr. \\
                                                      & |\delta_{l\tau} + C_{V_1}^l + C_{V_2}^l|^2 \left[ \left( 1 + {m_\tau^2 \over2q^2} \right) H_{V,0}^{s\,2} + {3 \over 2}{m_\tau^2 \over q^2} \, H_{V,t}^{s\,2} \right] \\
                                                      &+ {3 \over 2} |C_{S_1}^l + C_{S_2}^l|^2 \, H_S^{s\,2} + 8|C_T^l|^2 \left( 1+ {2m_\tau^2 \over q^2} \right) \, H_T^{s\,2} \\
                                                      &+ 3\Re[ ( \delta_{l\tau} + C_{V_1}^l + C_{V_2}^l ) (C_{S_1}^{l*} + C_{S_2}^{l*} ) ] {m_\tau \over \sqrt{q^2}} \, H_S^s H_{V,t}^s \\
                                                      &- 12\Re[ ( \delta_{l\tau} + C_{V_1}^l + C_{V_2}^l ) C_T^{l*} ] {m_\tau \over \sqrt{q^2}} \, H_T^s H_{V,0}^s \biggl.\biggr\} \,,
   \end{split}
\end{equation}
and

\begin{equation}
   \begin{split}
      & {d\Gamma(\Bbar \to \Dst \tau\nubar_l) \over dq^2} = {G_F^2 |V_{cb}|^2 \over 192\pi^3 m_B^3} q^2 \sqrt{\lambda_\Dst(q^2)} \left( 1 - {m_\tau^2 \over q^2} \right)^2 \times\biggl\{ \biggr. \\
      & \quad\quad\quad\quad\quad ( |\delta_{l\tau} + C_{V_1}^l|^2 + |C_{V_2}^l|^2 ) \left[ \left( 1 + {m_\tau^2 \over2q^2} \right) \left( H_{V,+}^2 + H_{V,-}^2 + H_{V,0}^2 \right) + {3 \over 2}{m_\tau^2 \over q^2} \, H_{V,t}^2 \right] \\
      & \quad\quad\quad\quad\quad - 2\Re[(\delta_{l\tau} + C_{V_1}^l) C_{V_2}^{l*}] \left[ \left( 1 + {m_\tau^2 \over 2q^2} \right) \left( H_{V,0}^2 + 2 H_{V,+} H_{V,-} \right) + {3 \over 2}{m_\tau^2 \over q^2} \, H_{V,t}^2 \right] \\
      & \quad\quad\quad\quad\quad + {3 \over 2} |C_{S_1}^l - C_{S_2}^l|^2 \, H_S^2 + 8|C_T^l|^2 \left( 1+ {2m_\tau^2 \over q^2} \right) \left( H_{T,+}^2 + H_{T,-}^2 + H_{T,0}^2  \right) \\
      & \quad\quad\quad\quad\quad + 3\Re[ ( \delta_{l\tau} + C_{V_1}^l - C_{V_2}^l ) (C_{S_1}^{l*} - C_{S_2}^{l*} ) ] {m_\tau \over \sqrt{q^2}} \, H_S H_{V,t} \\
      & \quad\quad\quad\quad\quad - 12\Re[ (\delta_{l\tau} + C_{V_1}^l) C_T^{l*} ] {m_\tau \over \sqrt{q^2}} \left( H_{T,0} H_{V,0} + H_{T,+} H_{V,+} - H_{T,-} H_{V,-} \right) \\
      & \quad\quad\quad\quad\quad + 12\Re[ C_{V_2}^l C_T^{l*} ] {m_\tau \over \sqrt{q^2}} \left( H_{T,0} H_{V,0} + H_{T,+} H_{V,-} - H_{T,-} H_{V,+} \right) \biggl.\biggr\} \,,
   \end{split}
\end{equation}
\end{appendices}
where $\lambda_{D^{(*)}}(q^2) = ((m_B - m_{D^{(*)}})^2 - q^2)((m_B + m_{D^{(*)}})^2 - q^2)$.

The $\tau$ polarization for a $B\to D^*$ decay ($P_\tau (D^*)$) is defined as:
     \begin{equation}
         P_\tau(\Dst) = { \Gamma(\lambda_\tau=1/2) - \Gamma(\lambda_\tau=-1/2) \over \Gamma(\lambda_\tau=1/2) + \Gamma(\lambda_\tau=-1/2) } \,,
         \label{eq:Ptau}
         \end{equation}
where:
\begin{subequations}
   \label{eq:GammaDst_tau}
   \begin{align}
      \label{eq:GammaDst_tau1}
      \begin{split}
         & {d\Gamma^{\lambda_\tau=1/2}(\Bbar \to \Dst\tau\nubar_l) \over dq^2} = {G_F^2 |V_{cb}|^2 \over 192\pi^3 m_B^3} q^2 \sqrt{\lambda_\Dst(q^2)} \left( 1 - {m_\tau^2 \over q^2} \right)^2 \times\biggl\{ \biggr. \\
         & \quad\quad\quad\quad\quad {1 \over 2} ( |\delta_{l\tau} + C_{V_1}^l|^2 + |C_{V_2}^l|^2 ) {m_\tau^2 \over q^2} \left( H_{V,+}^2 + H_{V,-}^2 + H_{V,0}^2 +3 H_{V,t}^2 \right) \\
         & \quad\quad\quad\quad\quad - \Re[ (\delta_{l\tau} + C_{V_1}^l) C_{V_2}^{l*} ] {m_\tau^2 \over q^2} \left( H_{V,0}^2 + 2 H_{V,+} H_{V,-} + 3H_{V,t}^2 \right) \\      
         & \quad\quad\quad\quad\quad + {3 \over 2} |C_{S_1}^l - C_{S_2}^l|^2 \, H_S^2 + 8|C_T^l|^2 \left( H_{T,+}^2 + H_{T,-}^2 + H_{T,0}^2  \right) \\
         & \quad\quad\quad\quad\quad + 3\Re[ ( \delta_{l\tau} + C_{V_1}^l - C_{V_2}^l ) (C_{S_1}^{l*} - C_{S_2}^{l*} ) ] {m_\tau \over \sqrt{q^2}} \, H_S H_{V,t} \\
         & \quad\quad\quad\quad\quad - 4\Re[ (\delta_{l\tau} + C_{V_1}^l) C_T^{l*} ] {m_\tau \over \sqrt{q^2}} \left( H_{T,0} H_{V,0} + H_{T,+} H_{V,+} - H_{T,-} H_{V,-} \right) \\
         & \quad\quad\quad\quad\quad + 4\Re[ C_{V_2}^l C_T^{l*} ] {m_\tau \over \sqrt{q^2}} \left( H_{T,0} H_{V,0} + H_{T,+} H_{V,-} - H_{T,-} H_{V,+} \right) \biggl.\biggr\} \,,
      \end{split} \\
      \nonumber \\
      \label{eq:GammaDst_tau2}
      \begin{split}
         & {d\Gamma^{\lambda_\tau=-1/2}(\Bbar \to\Dst\tau\nubar_l) \over dq^2} = {G_F^2 |V_{cb}|^2 \over 192\pi^3 m_B^3} q^2 \sqrt{\lambda_\Dst(q^2)} \left( 1 - {m_\tau^2 \over q^2} \right)^2 \times\biggl\{ \biggr. \\
         & \quad\quad\quad\quad\quad ( |\delta_{l\tau} + C_{V_1}^l|^2 + |C_{V_2}^l|^2 ) \left( H_{V,+}^2 + H_{V,-}^2 + H_{V,0}^2 \right) \\
         & \quad\quad\quad\quad\quad - 2\Re[ (\delta_{l\tau} + C_{V_1}^l) C_{V_2}^{l*} ] \left( H_{V,0}^2 + 2 H_{V,+} H_{V,-} \right) \\      
         & \quad\quad\quad\quad\quad + 16|C_T^l|^2 {m_\tau^2 \over q^2} \left( H_{T,+}^2 + H_{T,-}^2 + H_{T,0}^2 \right) \\
         & \quad\quad\quad\quad\quad - 8\Re[ (\delta_{l\tau} + C_{V_1}^l) C_T^{l*} ] {m_\tau \over \sqrt{q^2}} \left( H_{T,0} H_{V,0} + H_{T,+} H_{V,+} - H_{T,-} H_{V,-} \right) \\
         & \quad\quad\quad\quad\quad + 8\Re[ C_{V_2}^l C_T^{l*} ] {m_\tau \over \sqrt{q^2}} \left( H_{T,0} H_{V,0} + H_{T,+} H_{V,-} - H_{T,-} H_{V,+} \right) \biggl.\biggr\} \,.
      \end{split}
   \end{align}
\end{subequations}

The $D^*$ polarization $F_L(D^*)$ is defined as:
\begin{equation}
         F_L(D^*) = { \Gamma(\lambda_\Dst=0) \over \Gamma(\lambda_\Dst=0) + \Gamma(\lambda_\Dst=1) + \Gamma(\lambda_\Dst=-1) } \,.
         \label{eq:PDst}
      \end{equation}
where:
\begin{subequations}
   \label{eq:GammaDst_Dst}
   \begin{align}
      \label{eq:GammaDst_trDst}
      \begin{split}
         {d\Gamma^{\lambda_\Dst=\pm1}(\Bbar \to \Dst\tau\nubar_l) \over dq^2} =& {G_F^2 |V_{cb}|^2 \over 192\pi^3 m_B^3} q^2 \sqrt{\lambda_\Dst(q^2)} \left( 1 - {m_\tau^2 \over q^2} \right)^2 \times\biggl\{ \biggr. \\
                                                                               &  \left( 1 + {m_\tau^2 \over2q^2} \right) \bigl(\bigr. |\delta_{l\tau} + C_{V_1}^l|^2 H_{V,\pm}^2 + |C_{V_2}^l|^2 H_{V,\mp}^2 \\
                                                                               & \quad\quad\quad\quad\quad - 2\Re[ (\delta_{l\tau} + C_{V_1}^l) C_{V_2}^{l*}] H_{V,+} H_{V,-} \bigl.\bigr) \\
                                                                               &  + 8|C_T^l|^2 \left( 1+ {2m_\tau^2 \over q^2} \right) H_{T,\pm}^2 \\
                                                                               &  \mp 12\Re[ (\delta_{l\tau} + C_{V_1}^l) C_T^{l*} ] {m_\tau \over \sqrt{q^2}} H_{T,\pm} H_{V,\pm} \\
                                                                               &  \pm 12\Re[ C_{V_2}^l C_T^{l*} ] {m_\tau \over \sqrt{q^2}} H_{T,\pm} H_{V,\mp} \biggl.\biggr\} \,,
      \end{split} \\
      \nonumber \\
      \label{eq:GammaDst_longDst}
      \begin{split}
         {d\Gamma^{\lambda_\Dst=0}(\Bbar \to \Dst\tau\nubar_l) \over dq^2} =& {G_F^2 |V_{cb}|^2 \over 192\pi^3 m_B^3} q^2 \sqrt{\lambda_\Dst(q^2)} \left( 1 - {m_\tau^2 \over q^2} \right)^2 \times\biggl\{ \biggr. \\
                                                                            &  |\delta_{l\tau} + C_{V_1}^l - C_{V_2}^l|^2 \left[ \left( 1 + {m_\tau^2 \over2q^2} \right) H_{V,0}^2 + {3 \over 2}{m_\tau^2 \over q^2} \, H_{V,t}^2 \right] \\
                                                                            &  + {3 \over 2} |C_{S_1}^l - C_{S_2}^l|^2 \, H_S^2 + 8|C_T^l|^2 \left( 1+ {2m_\tau^2 \over q^2} \right) H_{T,0}^2 \\
                                                                            &  + 3\Re[ ( \delta_{l\tau} + C_{V_1}^l - C_{V_2}^l ) (C_{S_1}^{l*} - C_{S_2}^{l*} ) ] {m_\tau \over \sqrt{q^2}} \, H_S H_{V,t} \\
                                                                            &  - 12\Re[ ( \delta_{l\tau} + C_{V_1}^l - C_{V_2}^l) C_T^{l*} ] {m_\tau \over \sqrt{q^2}} H_{T,0} H_{V,0} \biggl.\biggr\} \,.
      \end{split}
   \end{align}
\end{subequations}

In all the above, the $H$'s are the so called helicity amplitudes which can be written down in terms of the corresponding hadronic form factor parameters. We refer to ref.~\cite{Sakaki:2013bfa} for a detailed description of the same. In accordance with the same reference, we have used the Caprini-Lellouch-Neubert (CLN)~\cite{Caprini:1997mu} parametrization for the $B\to D^{(*)}$ form factors.


\providecommand{\href}[2]{#2}\begingroup\raggedright\endgroup

\end{document}